# Self-organization of ultrasound in viscous fluids

V.J. Sánchez-Morcillo[1], J. Martínez-Mora[1], I. Pérez-Arjona[1], V. Espinosa[1], and P. Alonso[2]

[1]*Instituto de Investigación para la Gestión Integrada de las Zonas Costeras, Universidad Politécnica de Valencia, C/. Paranimf 1, 46730 Grao de Gandia, Spain*

[2] *Departamento de Sistemas Informáticos y Computación, Universidad Politécnica de Valencia, Cno. Vera s/n, 46022 Valencia, Spain*

We report the theoretical and experimental demonstration of pattern formation in acoustics. The system is an acoustic resonator containing a viscous fluid. When the system is driven by an external periodic force, the ultrasonic field inside the cavity experiences different pattern-forming instabilities leading to the emergence of periodic structures. The system is also shown to possess bistable regimes, in which localized states of the ultrasonic field develop. The thermal nonlinearity in the viscous fluid, together with the far-from-equilibrium conditions, are is the responsible of the observed effects.

Spontaneous pattern formation is a universal behavior observed in a diverse class of far from equilibrium systems in chemistry, biology, fluid mechanics, and nonlinear optics [1,2]. The latter has recently received great attention, owing to the potential applications in all-optical signal processing devices [2]. One noticeable exception in the list of disciplines where pattern formation has been demonstrated is acoustics. Although several nonlinear dynamic phenomena have been reported in acoustics, there not exists up to now a clear evidence of self-organization of sound, in spite of the strong similarities between the behaviour of light and sound waves. On one side, the observation of patterns requires systems with a large aspect ratio or Fresnel number, a situation which is not commonly considered in acoustics. On the other side, nonlinearity in typical fluids is quadratic and of elastic nature. This circumstance, together with the absence of dispersion of sound (but not optical) waves, results in higher harmonics generation with a complicated spectrum which makes difficult the observation and analysis of spatial instabilities. We present here what we believe to be the first experimental observation of spontaneous pattern formation of a confined ultrasonic field, propagating in a viscous medium and driven far from equilibrium.

The physical model we consider is an acoustic resonator or interferometer, formed by two flat, parallel and highly reflecting walls which contain a viscous fluid. One of the walls vibrates at ultrasonic frequencies $f$ with a given amplitude. For high enough driving amplitudes, the system behaviour is nonlinear, the main source of nonlinearity being of thermal (not elastic) origin. In a viscous medium the sound velocity $c$ depends significantly on temperature, $c = c_0(1 - \sigma T')$, where $c_0$ is the velocity of sound at some equilibrium (ambient) temperature, $T'$ denotes the variation of the medium temperature from that equilibrium due to the intense acoustic wave, and $\sigma$ is the parameter of thermal nonlinearity. The propagation of sound in such a medium has been shown to be described in terms of two coupled equations for pressure, $p'$, and temperature, $T'$, deviations [5]. A previous theoretical and experimental study in this system [6] evidenced the existence of nontrivial temporal dynamics (self-pulsations and chaos). All these studies were performed in the framework of the plane wave approximation, thus spatial coupling was ignored. This is a valid approximation when the system has a small aspect ratio (e.g. when the distance between the walls is large compared with the transverse size). We recently extended the plane wave theory by considering the spatial effects of sound diffraction and temperature diffusion in a large aperture resonator [7]. These effects, which are responsible of the spatial coupling, can play an important role when the Fresnel number of the resonator $F = l^2/\lambda L \gg 1$ (being $l$ and $L$ its transverse and longitudinal dimensions, respectively).

The dynamical model describing this system was presented in [7]. It is assumed that the pressure field is given by two counter-propagating waves, with slowly varying amplitudes in time and space. When the walls are highly reflecting one can consider the mean-field limit, where the amplitudes do not depend on the axial coordinate $z$, and the pressure wave is given by $p' = p(r_\perp,t)\cos(kz)\sin(\omega t)$. The temperature field is decomposed into a homogeneous and a grating component, and takes the form $T = T_h(r_\perp,t) + T_g(r_\perp,t)\cos(2kz)$.

In terms of the normalized set of variables $P = p(\sigma \omega t_p t_h \alpha_0 / 2\rho_0^2 c_0 c_p)^{1/2}$, $H = \omega t_p \sigma T_h$,

$G = \omega t_g \sigma T_g$, the evolution of the system is determined by the dynamical equations [7]

$$\tau_p \frac{\partial P}{\partial \tau} = -(1+i\Delta)P + P_{in} + i\nabla^2 P + i(H+G)P$$
$$\frac{\partial H}{\partial \tau} = -H + D\nabla^2 H + 2|P|^2 \qquad (1)$$
$$\frac{\partial G}{\partial \tau} = -\tau_g^{-1} G + D\nabla^2 G + |P|^2$$

where $\tau = t/t_h$ is the time measured in units of the relaxation time $t_h$ of the homogeneous temperature component, and $\tau_p = t_p/t_h$ and $\tau_g = t_g/t_h$ are the normalized relaxation times of the intracavity pressure field and the temperature grating component, respectively. Their original values are given by $t_p^{-1} = c_0 T/2L + c_0 \alpha_0$, where $T$ is the transmission coefficient of the plates and $\alpha_0$ is the absorption coefficient of the medium, and $t_g^{-1} = 4\kappa^2 \chi$, where $\chi = \kappa/\rho_0 c_p$ is the coefficient of thermal diffusivity, $\rho_0$ is the equilibrium density of the medium and $\kappa$ and $c_p$ are the thermal conductivity and the specific heat of the fluid at constant pressure, respectively. Other parameters are the detuning $\Delta = (\omega_c - \omega)t_p$, with $\omega_c$ the cavity frequency that lies nearest to the driving frequency $\omega$, and $P_{in} = c_0 F p_{in}/2L$, being $p_{in}$ the injected pressure plane wave amplitude, which we take as real without loss of generality. Finally, in the transverse Laplacian operator the dimensionless transverse coordinates $(x,y)$ are measured in units of the diffraction length $l_d = c_0 (t_p/2\omega)^{1/2}$, and the normalized diffusion coefficient $D = \chi\, t_h /l_d^2$. We note that when spatial derivatives are ignored the model of [6] is retrieved.

The model parameters can be estimated for a typical experimental situation. We consider a resonator with high quality plates (T = 0.1), separated by a distance $L$, driven at a frequency $f$ = 2 MHz, and containing glycerine at 10°C. Under these conditions the medium parameters are $c_0$=2×10$^3$ m/s, $\alpha_0$=10 m$^{-1}$, $\rho_0$=1.2×10$^3$ kg/m$^3$, $c_p$ = 4×10$^3$ J/kg K, $\sigma = 10^{-2}$K$^{-1}$, and $\kappa$ = 0.5 W m$^{-1}$K$^{-1}$ ($\chi$ =10$^{-7}$ m$^2$s$^{-1}$). In this case $t_p$=2×10$^{-5}$ s, $t_g$ = 6×10$^{-2}$ s, and our length unit $l_d$ = 2 mm. The Fresnel number of the system is 40. For a resonator with a large Fresnel number, the relaxation of the homogeneous component of the temperature is mainly due to the heat flux through the boundaries, and can be estimated from the Newton's cooling law as $t_h$ = 10$^1$ s, which is used as the time unit. Then the diffusion constant D = 10$^0$, and the normalized decay times $\tau_p$ = 10$^{-6}$, and $\tau_g$ = 10$^{-2}$ under usual conditions. Note that the problem is typically very stiff, the decay rates obeying $0 < \tau_p \ll \tau_g \ll 1$.

The spatially uniform steady state is obtained by neglecting the derivatives in Eqs. (1). It is convenient to introduce the new magnitudes $\overline{X} = |P|^2$ and $\overline{Y} = |P_{in}|^2$, which are proportional to the intensities. The stationary temperature fields are readily obtained as $\overline{H} = 2\overline{X}$ and $\overline{G} = \tau_g^{-1} \overline{X}$, where the stationary pressure can be found from the nonlinear equation

$$\overline{Y} = \overline{X} + \left(\Delta - 2\overline{X}^2\right)\overline{X}. \qquad (2)$$

The characteristic curve $\overline{X}$ vs $\overline{Y}$ can display an S-shape when $\Delta > \sqrt{3}$, as shown in the inset in Fig. 1.

The existence of pattern forming instabilities of the solutions of Eq. (2) is explored by performing a linear stability analysis against space-dependent perturbations in the form $e^{\lambda(k_\perp)t}e^{i\mathbf{k}_\perp \mathbf{r}_\perp}$, where $\lambda(k_\perp)$ represents the growth rate of the perturbations, and $k_\perp$ is the transverse component of the wavevector, responsible for the appearance of modulations in the amplitude profile.

The instability domain can be numerically evaluated in the parameter space defined by $<\Delta,P_{in}>$, if we fix the rest of parameters ($D$, $\tau_p$ and $\tau_g$). This is a convenient representation, since detuning and injection are the single parameters which can be varied in the experiment, and the other parameters are mainly determined by the fluid properties. The result is shown in the Figure 1 at the right.

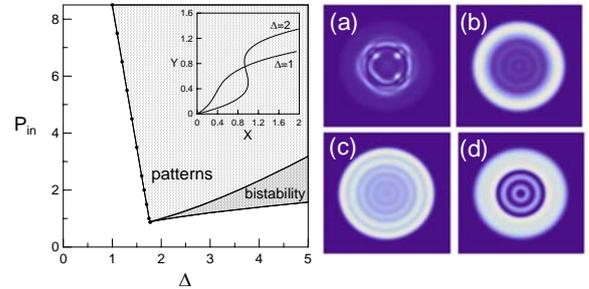

*Figure 1. Right: Instability domain. The inset shows the monostable and bistable solutions of Eq. (2). Left: spatial structures obtained by numerical integration of Eqs. (1).*

The dotted and shaded regions correspond to parameter values resulting in modulationally unstable solutions. Note that the region of bistable solutions, existing for $\Delta > \sqrt{3}$, is also modulationally unstable. This result was discussed in [7] in the framework of a generalized Swift-Hohenberg equation, which is a simplified description (an order parameter equation) of Eqs. (1) valid in the vicinity of the nascent bistability point. In this region, where modulationally stable and unstable solutions coexist, the systems support stable localized structures or cavity solitons, as demonstrated in [7].

With the aim of checking the predictions of the linear stability analysis, we integrated numerically Eqs. (1) by using the split-step technique on a spatial grid of dimensions 128X128. In order to mimic the experimental conditions produced by the exciting transducer, with the amplitude vanishing at the boundaries, a hypergaussian distribution with radial symmetry (top hat profile) was used for the pump term. The parameters in the simulations are such that a pattern forming instability is predicted [corresponding to the shaded region in Fig. (1)].

We note that, as discussed before, the simulated model is stiff (the decay rates of the different fields differ by orders of magnitude), which makes the calculations extremely time and memory consuming. For this reason, the simulations were performed on a Graphical Processor Unit (GPU) allowing for parallel computing. Some typical results of the numerical simulations are shown in Fig. 1. In Fig. 1(a) parameters in the bistable region ($P_{in}$ = 1.5, $\Delta$=3) lead to several localized structures, while in Figs 1.(b)-(d) the system evolve to stable homogeneous solutions [Figs. 1(b) and (c)] or to modulated patterns [Fig. 1(d)] for $P_{in}$ = 5 and $\Delta$ = –2, 0 and 2 respectively. Note that the weak modulations in cases (b) and (c) are the result of boundary effects, and disappear as the area of the integration domain increases. Other parameters are D = 1, $\tau_p = 10^{-6}$ and $\tau_g = 10^{-2}$.

An experimental setup was designed to check the theoretical predictions. The resonator consists in two piezoceramic discs with radius 1.5 cm mounted in a Plexiglas tank containing glycerin. Both sides are located at a variable distance L, and its parallelism can be carefully adjusted to get a high-Q interferometer. One of the piezoceramics, with a resonance frequency around 2 MHz, is driven by the signal provided by a function generator (Agilent 33220) and a broadband RF power amplifier ENI 240L. The experimental setup is completed by a needle hydrophone (TNUA200 NTR Systems) to measure the intracavity pressure field. The hydrophone has a sensitive section of 0.2 mm and is mounted on an OWIS GmbH two-axis motorized system that allows scanning the transversal XY plane with a repeatability of 0.25 mm. In this way, we measured the transverse spatial distribution of the pressure field at a fixed axial position, close to the emitter transducer. The tank, with dimensions 25X25X25 cm$^3$, is made of Plexiglas, and covered with an absorbent material to avoid reflections at the tank walls. A cooling system is used to keep temperature variations from the environment as stable as possible.

A preliminary study of the cavity resonances was done, in order to find the experimental values of the frequency detuning (for a given input) and the cavity linewidth $\gamma = t_p^{-1}$. These parameters allow to evaluate the theoretical amplitudes given by Eq. (2), which will be compared with the experimental results. Figure 2 shows the dependence of the intracavity amplitudes with the injection (bifurcation diagrams) for a cavity length $L = 2$ cm, and a source frequency of $f = 1.75$ MHz (a) and $f = 2.03$ MHz (b). The temperature of the tank was kept constant at $T = 24$ °C. The cavity detuning is obtained as $\Delta = (\omega - \omega_c)/\gamma \approx 1.6$ in the case (a) and $\Delta \approx 2$ in the case (b), where $\omega_c$ is the closest cavity eigenmode. Note that these values of detuning correspond to the monostable and bistable regimes of the stationary solutions, being below (a) and above (b) the critical point of nascent bistability $\Delta_c \approx 1.73$.

In Fig. 2(b) the experimental results are presented with different symbols. The pump value is increased from 0.07 to 0.25 V, and then decreased back to the initial value. Black dots represent the measured values when the driving amplitude approaches to the hysteretic region, from below (left branch) or from above (right branch)

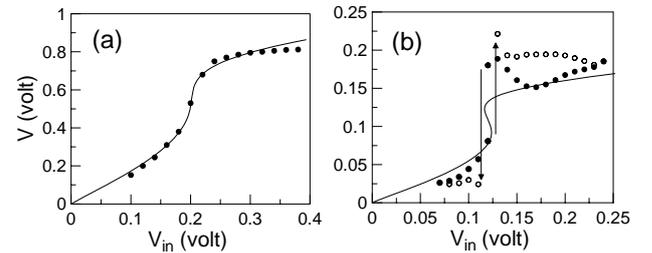

*Figure 2. Bifurcation diagram in monostable (a) and bistable (b) cases. Dots correspond to the experimental results, and the solid line to the theoretical prediction for f=1.75 MHz (a) and for f=2.03 MHz. Detuning values used in Eq. (2) are obtained from the experiment as $\Delta$=1.6 and $\Delta$=2.0 respectively*

Note that the measured amplitudes in the upper branch (with higher intensities) are different, depending on the direction of the pump variation. Also, the measured values fit quite well to the theoretical prediction *except* in the vicinity of the jump between branches of stationary solutions. We believe that these discrepancies, not observed in the monostable regime in Fig. 2(a), can be attributed to the finite relaxation time of the temperature field. After the transition from one branch to the other, the local properties of the medium are strongly affected, and the system needs a finite time to reach again the equilibrium state. This time is larger than the interval between two consecutive measurements. Note that, as we move along a given branch, the solutions smoothly tend to their stationary values.

A scanning of the transverse section of the resonator evidences the emergence of spatial structures similar to those predicted by the numerical simulation in Fig. 1. The parameters in Fig. 3 are those corresponding to Fig. 2(a), i.e. we explore here pattern formation in the monostable regime. In Figs. 3(a) and (b) the measured amplitude and its spatial spectrum are shown, as obtained for an input amplitude $V_{in}$=0.4 mV. Here the basic structure is a target pattern, slightly modulated in the azimutal direction (the central part of the beam is shown). A similar representation is shown in Figs. 3 (c) and (d), but for a higher input $V_{in}$=0.9 mV. Note that the azimutal symmetry is broken, and a flower-like pattern develops. The spatial spectrum in Figs. 3(b) and (d) indicate the occurrence of the instability: the ring-like spectrum is transformed into a discrete spectrum, where wavevectors at particular angles dominate.

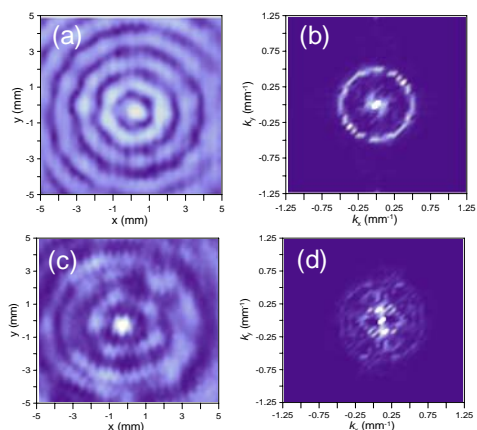

*Figure 3. Experimental patterns observed for Δ = 1.6. (a) and (c) show spatial distributions, while (b) and (d) the spatial spectrum.*

The existence of localized structures (LSs) has been also evidenced by the experiment. According to the theory, they exist for parameter values corresponding to the hysteretic region in Fig. 2(b). Figure 4(a) shows an experimentally observed LS, which emerged from a noisy homogeneous background amplitude at an arbitrary (non-centered) position of the transverse plane. The numerical simulation result is shown in Fig. 4(b), with the amplitude normalized to the peak value. In both cases, the size if the LS is 3-4 mm. Note also the agreement between the amplitudes of the peaks and the background field. In order to compare with measurements, in Fig. 4(b) pressure was converted to voltage using the calibration factor of the hydrophone, 1.26 V/MPa at $f$ = 2 MHz.

In conclusion, we have investigated both theoretical and experimentally the transverse dynamics of the acoustic field inside a driven resonator containing a viscous fluid. The results demonstrate, for the first time, the existence of different self-organization scenarios depending on the parameters of the system. Both extended patterns (target- and flower-like) and localized structures (isolated peaks on a homogeneous background) have been observed, in agreement with the theoretical predictions. Let us finally mention that the results reported here present strong similarities with those reported for vertical cavity semiconductor lasers (VCSELs) in [8].

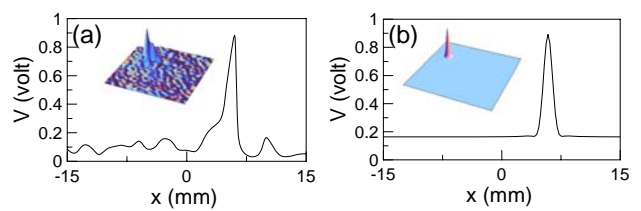

*Figure 4. Localized structure, obtained experimentally (left) and theoretically (right) for Δ = 2. A cross-section on the LS is shown, together with the complete spatial profile (inset).*

The work was financially supported by Spanish Ministerio de Ciencia e Innovación and European Union FEDER through project FIS2008-06024-C03.

___